\documentclass[preprint2]{aastex}
\shorttitle{Magnetic vector fields of an AR filament}
\shortauthors{XU et al.}
\begin{document}
\title{Magnetic fields of an active region filament from full Stokes analysis of Si I 1082.7 nm and He I 1083.0 nm}
\author{Z. Xu \altaffilmark{1},A. Lagg \altaffilmark{2}, S. Solanki\altaffilmark{2,3},Y. Liu \altaffilmark{1}}
\email{xuzhi@ynao.ac.cn}
\altaffiltext{1}{Yunnan Astronomical Observatory/National Astronomical Observatories, Chinese Academy of Science, Kunming 650011, PR China.}
\altaffiltext{2}{Max-Planck-Institut f\"ur Sonnensystemforschung, Max-Planck-Strasse 2, 37191  Katlenburg-Lindau, Germany}
\altaffiltext{3}{School of Space Research, Kyung Hee University, Yongin, Gyeonggi 446-701, Republic of Korea}
\begin{abstract}
Vector magnetic fields of an active region filament in the photosphere and upper chromosphere are obtained from spectro-polarimetric observations recorded with the Tenerife Infrared Polarimeter (TIP II) at the German Vacuum Tower Telescope (VTT). We apply Milne-Eddington inversions on full Stokes vectors of the photospheric Si I 1082.7 nm and the upper chromospheric He I triplet at 1083.0 nm to obtain magnetic field vector and velocity maps in two atmosphere layers. We find that: (1)A complete filament was already present in H$\alpha$ at the beginning of the TIP II data acquisition. Only a partially formed one, composed of multiple small threads, was present in He I. (2) The AR filament comprises two sections. One shows strong magnetic field intensities, about 600 - 800 G in the upper chromosphere and 800 - 1000 G in the photosphere. The other exhibits only comparatively weak magnetic field strengths in both layers. (3) The Stokes V signal is indicative of a dip in the magnetic field strength close to the chromospheric PIL. (3) In the chromosphere consistent upflows are found along the PIL flanked by downflows. (4) The transversal magnetic field is nearly parallel to the PIL in the photosphere and inclined by 20$^\circ$ - 30$^\circ$ in the chromosphere. (5) The chromospheric magnetic field around the filament is found to be in normal configuration, while the photospheric field presents a concave magnetic topology. The observations are consistent with the emergence of a flux rope with a subsequent formation of a filament.

\end{abstract}


\section{Introduction}
Active region filaments, so named due to their location in or close to active regions, are generally short-lived, low-lying in the atmosphere and easily observed on the solar disk. Of particular interest are their obvious helical magnetic structures and stronger magnetic field strengths as compared to quiescent
filaments \citep[see the introduction of][]{harvey06,mackay10}. Helical structures are frequently observed in erupting filaments (prominences). One of the grandest
examples was recorded on 4 June 1946 at the Climax station of the High Altitude Observatory \citep[e.g.][]{anzer70}. For AR filaments, many observations
have revealed the presence of magnetic flux ropes in filament channels either by using vector magnetic fields retrieved from photospheric lines showing dips in filament
channels  \citep[e.g.][]{lit95,lit05,lopez06}, or by using nonlinear force-free extrapolations which exhibit twisted flux ropes \citep[e.g.][]{canou10, Guo10}.
These studies were based on the measurement of magnetic vectors at one level in the atmosphere (i.e. the photosphere). Many measurements of the magnetic field vector \emph{within} filaments have also been made using either the Zeeman or Hanle effect \citep{rust67,landi82,leroy83,leroy84,bom94,bom05,lopez02}. Most of the measurements were carried out for quiescent prominences and used the Stokes profiles of the D$_{3}$ line of Helium (see \cite{paletou03} for a review). An interesting result reported by \cite{bom94} shows that the majority of quiescent prominences (12 of the 14 samples) are of the inverse polarity configuration (IP, i.e. the direction of the horizontal component of the field inside the filament is opposite to the direction determined by the polarity at the photosphere). Only 2 of the 14 cases, both lying in the vicinity of an active region, are found in the normal polarity configuration (NP, i.e. the horizontal component of the field vectors are pointing in the sense suggested by the adjacent photospheric polarities).

In this work, we investigate the magnetic field configuration within an AR filament, using another important spectral window around the He I 1083.0 nm line, which has become an important tool to determine the magnetic field vector in the upper chromosphere \citep[e.g.][]{ruedi95,ruedi96,lin98,truj02,solanki03,lagg04,merenda06,xu10}. It also allows, through the simultaneously recorded Si I 1082.7nm line, both photospheric and chromospheric field vectors to be measured in a straightforward manner. Up to now spectro-polarimetric measurements of AR filaments using full Stokes vectors of the He I 1083.0 nm line are rare. Observations by \cite{sasso07} determined for the first time the magnetic field structure of an AR filament during its eruption phase and revealed the presence of different unresolved atmospheric components coexisting within one resolution element ($\sim$1.2 $\arcsec$). The magnetic field strengths they retrieved using a Milne-Eddington based inversion are in the range of 100 - 250 G \citep{sasso11}. Another measurement of magnetic field vectors in an AR filament was reported by \cite{kuckein09}. This filament lay over a plage region and was stable during the observation. These authors applied three independent inversion methods --- the weak-field approximation, a Milne-Eddington based inversion and a Principal Component Analysis (PCA) inversion including atomic polarization, on the full Stokes vector of the He I 1083.0 nm line. All three methods consistently provided field strengths in the range of 600 - 700 G at the formation height of the He I line, the highest strengths reported in filaments so far. Hence AR filaments appear to possess much stronger fields than quiescent prominences, which exhibit magnetic field strengths of the order of a few Gauss \citep[e.g.][]{truj02,merenda06}, albeit when observed above the solar limb, so that a part of the difference may be due to differences in the height sampled.

In this work, we present another observation of an AR filament lying over a region with multiple sunspots. We analyze and compare simultaneously recorded full Stokes
vector measurements of the photospheric Si I 1082.7 nm and chromospheric He I 1083.0 nm triplet. This allows us to study the magnetic field of the filament co-spatially
and co-temporally in the photosphere and at the formation height of the He I line. This is the same filament as studied by \cite{sasso07,sasso11}, but observed almost one day prior to its eruption, so that here we concentrate on its stable phase. Using the data taken in this phase, it should is more straightforward to determine the vector magnetic fields in both atmospheric layers, which allows us not only to test whether the strong magnetic field strengths found by \cite{kuckein09} are also present in other AR filaments, but also to investigate the three dimensional magnetic field structures around AR filaments, including the direct measurement of the magnetic field twist at multiple
heights \citep[e.g.][]{merenda06,asensio10}.

\section{Observations}
\begin{figure}
\centering
\includegraphics[width= 7 cm]{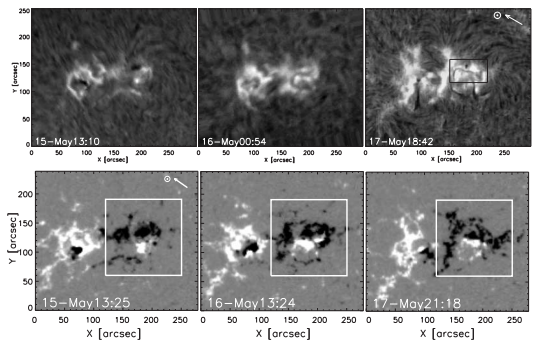}
\caption{Upper panels: H$\alpha$ images of the active region NOAA 10763 on 2005 May 15, 16 and 17 (taken from the full-disk H$\alpha$ images recorded by Kanzelh\"ohe Solar Observatory, Yunnan Observatory and Big Bear Solar Observatory, respectively). The images have almost the same field of view size. The area scanned by the VTT on May 17 is outlined by a box in the right frame. The arrow in the right frame points to solar disk center. Lower panels: Sequence of \emph{SOHO/MDI} magnetograms of the active region NOAA 10763 covering the period from May 15 to 17. A white box indicates a subregion of 130$\arcsec$ $\times$ 130$\arcsec$, which contains the studied filament and in which the magnetic flux evolution is measured. The direction of solar disk center is indicated by an arrow.
}
\label{fig1}
\end{figure}

Several filaments became visible in H$\alpha$ in active region NOAA 10763 from 2005 May 15 to 17 during which time it transited the solar meridian. Observations made by the Global H$\alpha$ Network run by Big Bear Observatory, shown in Fig.~\ref{fig1} (top row), reveal that the filament investigated in this paper (in the boxed area of the upper-right frame) began to be present in H$\alpha$ on May 16. The evolution of the magnetic field of the active region during this period is also plotted in Fig.~\ref{fig1} (bottom row) showing magnetograms recorded by SOHO/Michelson Doppler Imager \citep[MDI,][]{scherrer95}. The filament lay along the roughly E-W directed polarity inversion line (PIL). The two opposite polarities in the region of the filament converged towards each other during May 15 to 17. We estimate the magnetic flux in an area of 130$\arcsec$ $\times$ 130$\arcsec$ (outlined by a white box) by integrating separately the positive and negative polarity fluxes and investigate their temporal evolution during the three days. It is found that both the negative and positive flux first increased, arriving at a maximum on May 16, which is when the analyzed filament began to be visible in the H$\alpha$ line, then decreased at a nearly constant rate. In addition, we found that the convergence motion of opposite polarities took place mainly during the time when the magnetic flux is decreasing. Although the amount of positive and negative flux in the box differs by a factor of roughly 3 - 8, the decrease in the flux is almost the same and equal to approximately $1.5*10^{21}$ Mx (there was also a similar, but not exactly equal, increase in flux in both polarities on May 15). This decrease of equal amounts of positive and negative LOS magnetic fluxes suggests that the flux changes inside the boxed area are related to the convergence of opposite polarities, rather than due to fluxes from elsewhere moving into the box and later moving out of the box. We checked this hypothesis and confirmed that indeed significantly less flux passed through the boundary of the box than the flux change seen inside the box.

\begin{figure}
\centering
\includegraphics[width= 8 cm]{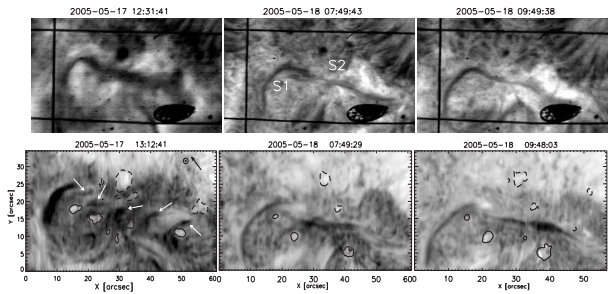}
\caption{Filament morphologies in the H$\alpha$ and He I 1083.0 nm line cores on May 17 and 18 observed by the VTT. Upper panels: slit-jaw images in H$\alpha$ line core at the times listed at the top of each image. The position of the slit at that time is visible as a vertical line. The field of view along the slit is bounded by the two horizontal lines.
Samples S1 and S2 in the middle panel indicate the main sections of the H$\alpha$ filament body (see the text). The large oval black spot at the bottom-right corner and other small dark spots are due to blemishes on the slit-jaw mirror. Lower panels: the intensity ratio between the He I line core and the continuum, obtained from scans made by the TIP II instrument. The onset time of each scan is given at the top of the corresponding image. Absorption features seen in the He I line are
indicated by white arrows in the lower left panel. Sunspots positions and polarities are outlined by contours. Solid (dashed) contour lines signify positive (negative) polarity sunspots. The direction of the disk center is indicated by a black arrow.
}
\label{fig2}%
\end{figure}

From May 17 to 18 this filament experienced several eruptions and flares. Observations made at the German Vacuum Tower Telescope (VTT) on Tenerife, both the slit-jaw recordings in the H$\alpha$ line and the spatial scans with the He I line, clearly show the variation of the filament morphology during these two days. We took one scan on May 17 and three scans on May 18. The active region was located close to the solar disk center on these days, at S16 W09 ($\mu$ = 0.96) on May 17 and at S14 W24 ($\mu$=0.94) on May 18 (here $\mu$ = cos$\theta$, $\theta$ is the heliocentric angle, i.e. the angle between the local solar surface normal direction and the line-of-sight direction). In Fig.~\ref{fig2} we plot the slit-jaw images (top row), which were taken during good seeing conditions and close to the onset time of each scan. These can be compared with He I line core images (bottom row) produced from the scans. The images in the He I line actually display the intensity ratio between the He I line core and the continuum in order to highlight the absorption features in the chromosphere. The third scan on May 18 is not shown in this figure since there was a major flare taking place at that time, which activated the filament and changed its morphology \citep[see the work of][]{sasso07}. In general, the H$\alpha$ filament body contains two main sections (identified in the middle panel of the top row): One section that lies nearly along the slit (S1) and another section nearly perpendicular to the slit (S2). Section S1 has very similar morphology in both H$\alpha$ and the He I lines and changes only little over the two days. On the contrary, S2 shows a markedly different morphology in the H$\alpha$ and He I lines. Particularly, on May 17, there are only several discrete absorption patches present in the filament channel in the He I image (indicated by arrows), which are unlike the complete and elongated absorption observed in H$\alpha$. The absorption patches seen in the He I line, running over the filament's spine (long axis of the filament) are around 2$\arcsec$ - 3$\arcsec$ wide and 7$\arcsec$ - 8$\arcsec$ long, and are separated from each other by 5$\arcsec$ - 10$\arcsec$. With time, these dark patches in He I grow and probably merge until they gradually resemble the filament seen in H$\alpha$.

Because a flare took place in the filament area between May 17 and May 18, which could have affected the subsequent evolution of the filament, we concentrate on analyzing the observations on May 17, when the filament was in its early stable phase and do not consider the observations on May 18 further. $\footnote{The flare mentioned here took place at 23:47 UT on May 17, i.e. 15 hours before the one studied by \cite{sasso11}.}$

\begin{figure}
\centering
\includegraphics[width= 7 cm]{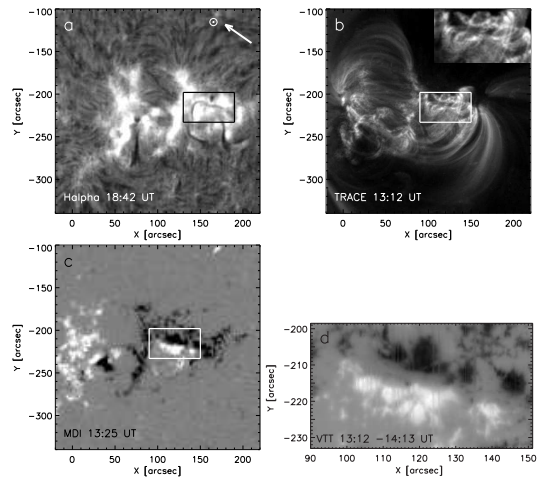}
\caption{\textbf{a}: H$\alpha$ image of the active region NOAA 10763 on 2005 May 17, taken from the full-disk H$\alpha$ image from Big Bear Solar Observatory (already shown in Fig.~\ref{fig1}). The direction of solar disk center is indicated by an arrow. \textbf{b}: EUV image of the active region
taken by \emph{TRACE} at 171 \AA. The filament (boxed area) is enlarged and displayed in the upper-right corner of this panel. \textbf{c}: Magnetogram of the active
region obtained by \emph{MDI}. \textbf{d}: Si I 1082.7 nm Stokes-$V$ map of the part of the active region scanned by the TIP II instrument.
All the panels are in the heliocentric disk coordinates. The boxes in panels a, b, and c identify the field of view of panel d. The observation time of each image is given in the lower left of each panel. For TIP II it is a range of times taken to scan the region.}
\label{fig3}
\end{figure}

In Fig.~\ref{fig3}, we provide more context images obtained on May 17. These include the H$\alpha$ image (repeated from the upper-right frame of Fig.~\ref{fig1}), a cutout from a 171 \AA\ channel image made by \emph{TRACE} spacecraft, a magnetogram taken by \emph{SOHO/MDI} and a Stokes-$V$ map obtained from the analyzed spectro-polarimetric scan made by the VTT (see below), respectively.

Spectro-polarimetric observations were carried out over a region containing the studied filaments using the TIP II polarimeter \citep{Collados07} installed on VTT. We set the slit (0.5$\arcsec$ wide and 35$\arcsec$ long) almost perpendicular to the H$\alpha$ filament and carried out spatial scans with a 0.17$\arcsec$ step size. At each scan position, full Stokes vectors were recorded over a spectral range of 11 \AA\, containing the photospheric Si I 1082.7 nm line and the chromospheric He I 1083.0 nm triplet. The exposure time per scan position was 10 seconds, resulting in a noise level of typically 5$\times10^{-4}$$I_{c}$ (where $I_{c}$ is the continuum intensity). The rectangles in Figs.~\ref{fig3}a, b and c delimit the part of the active region that was scanned with the TIP II instrument. In Fig.~\ref{fig3}d, a Stokes-$V$ map of the area scanned by TIP II is displayed in the same system of coordinates as the other panels (with solar disk center at the origin). The map is obtained by integrating the Stokes $V/I_{c}$ parameter within a wavelength range of 0.4 \AA\ in the blue wing of the photospheric Si I line. In this figure, we can clearly distinguish the studied AR filament both in H$\alpha$ and at 171 \AA. It lies almost exactly above the PIL separating two opposite polarity magnetic flux patches. Bright coronal arcades are found overlying the filament in the \emph{TRACE} image (The boxed area is enlarged and displayed in the upper-right corner of panel b).

\section{Retrieval of Stokes Profiles}
The spectro-polarimetric scans on May 17 covering the analyzed filament, were reduced by applying standard TIP data reduction procedures to the observed
Stokes profiles, including dark current subtraction, flat-fielding and polarimetric calibration involving a cross-talk removing algorithm \citep{beck05}.
An accurate continuum correction was carried out by comparing the average flat field profile with an FTS spectrum \citep{Delbouille81} of the average quiet sun.
The quiet-sun Stokes $I$ profile, computed by averaging the 100 profiles with the lowest polarization signal in the observed map, was used to determine the wavelength
calibration by assuming that the core position of the photosphere spectral lines correspond to the laboratory wavelength minus 200 m/s (corresponding to the granular blueshift). In addition, we applied a 5-pixel binning along the wavelength and 2-pixel binning along the slit direction. The estimated spatial resolution of the image made from each slit scan was limited by the seeing to
roughly 1$\arcsec$-2$\arcsec$.

Then, we applied an inversion code based on the Milne-Eddington approximation, HeLIx$^{+}$ \citep{lagg04, lagg09}, to the full Stokes vectors of the Si I and He I lines, normalized to $I_{c}$, to obtain the magnetic field vectors in the photosphere and upper chromosphere. The Si I line is inverted independently of the He I triplet. The HeLIX$^{+}$ code takes into account the incomplete Paschen-Back effect in the He I triplet \citep{Socas-Navarro04,sasso06} and obtains the best fit to the observed profiles by varying eight free parameters for a given atmospheric component: the magnetic field strength ($B$), its inclination relative to the LOS ($\gamma$) and azimuth angle ($\chi$), line-of-sight velocity ($v_{\rm los}$), Doppler width ($\Delta$$\lambda_{D}$), damping constant ($a$), slope of the source function ($S_{1}$) and the opacity ratio between line-center and continuum ($\eta_{0}$). An additional free parameter, the filling factor, $f$, is used when more than one atmospheric component is considered simultaneously.

\begin{figure}
\plottwo{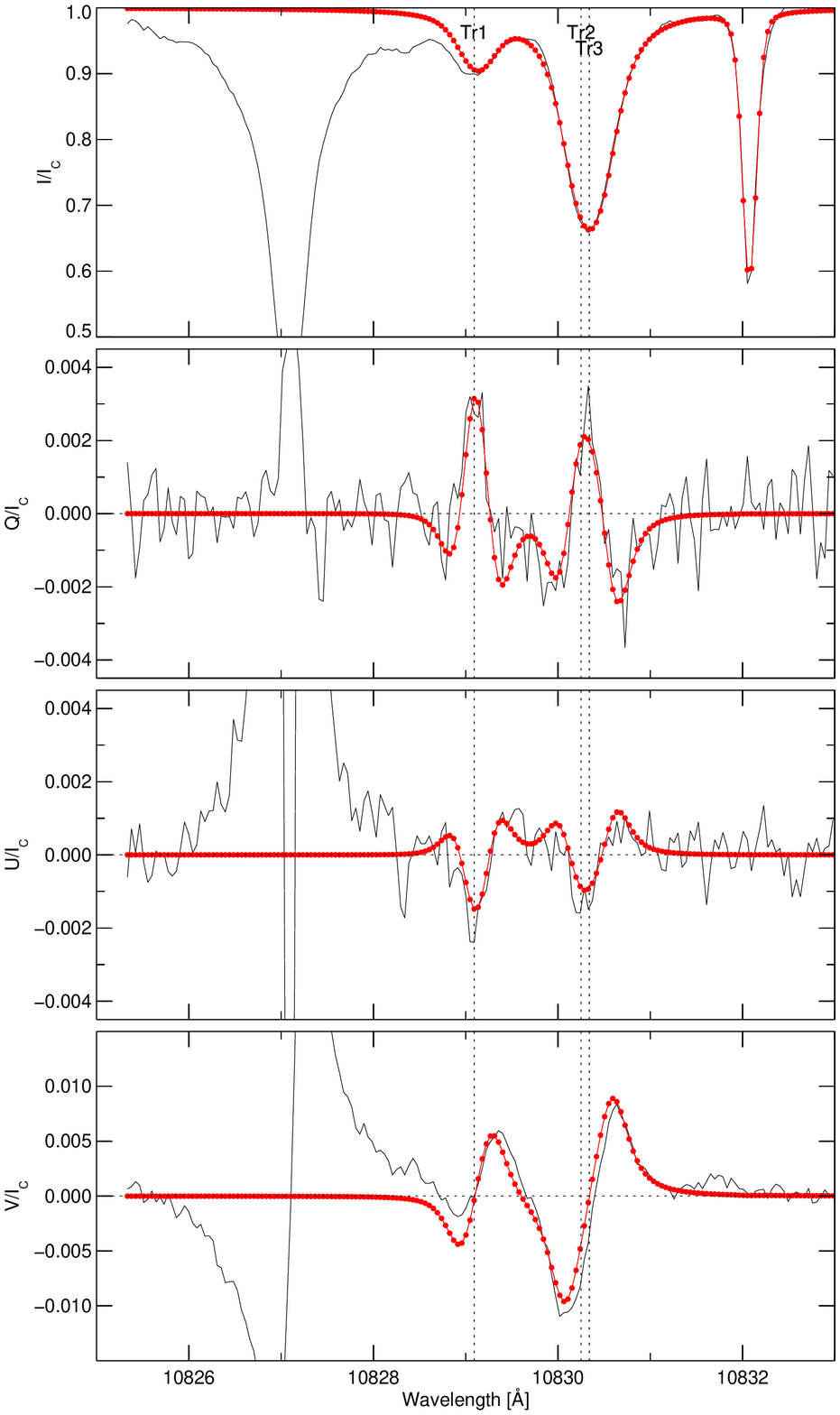}{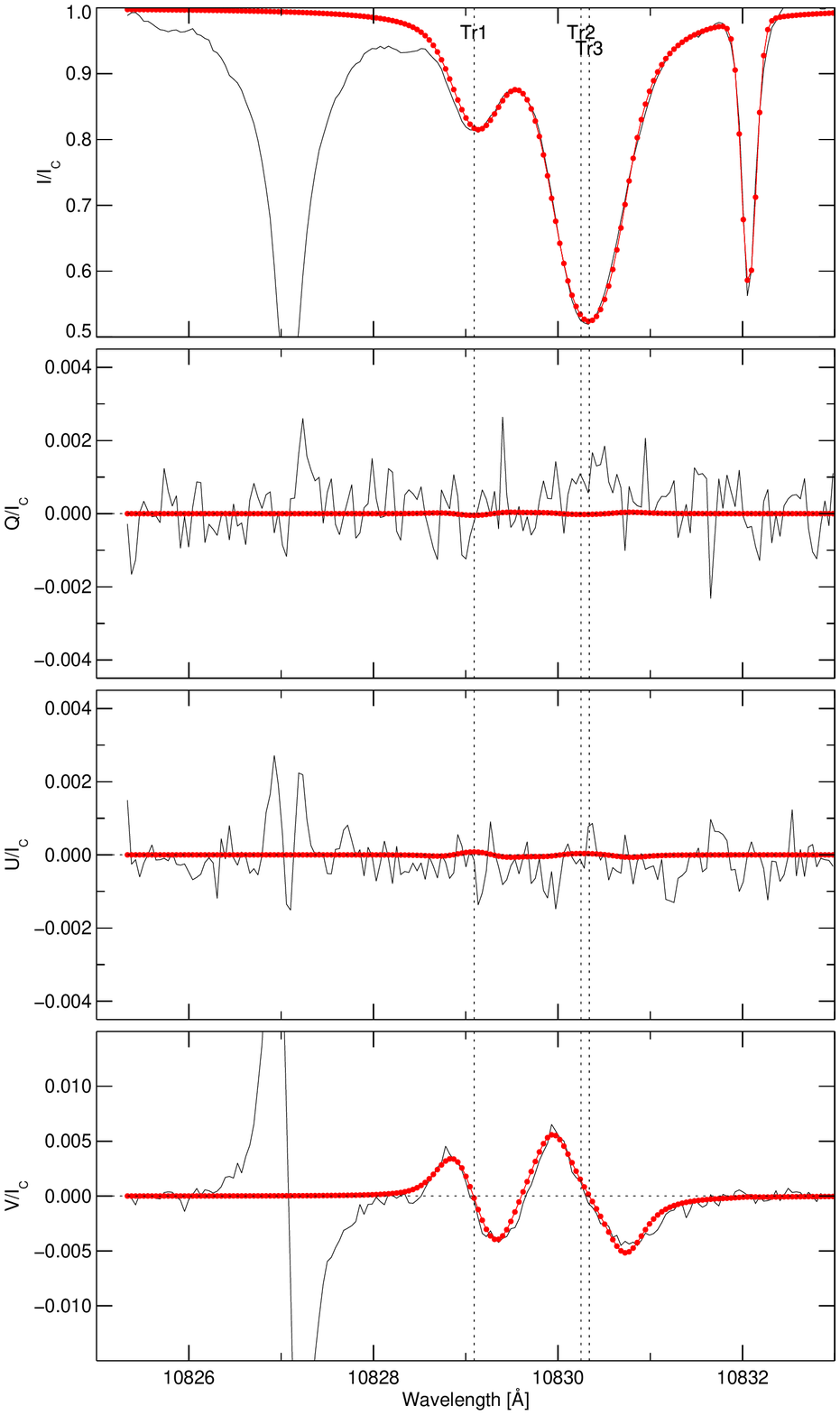}
\caption{Typical Stokes profiles of the He I triplet in the filament. \emph{Left}: Stokes profiles taken from one of the dark absorption patches
in section S2 of the filament at X = 22, Y = 19 in the coordinates of the lower-left panel of Fig.~\ref{fig4}. The Stokes $Q$ and $U$ profiles are clearly influenced by the transverse Zeeman effect, but do not show the signature of the Hanle effect. The observed profile is shown in black (solid line). The best fit by a Milne-Eddington inversion is shown in red (filled circles). Three vertical dotted lines indicate the line-center rest positions of the components of the He I triplet. The magnetic field
retrieved is $B$ =751 G, $\gamma$ = 105$^{\circ}$ and $\chi$ = 77$^{\circ}$. \emph{Right}: Stokes profiles taken from the filament section S1 at X =10, Y =17. The retrieved magnetic field is $B$ =164 G, $\gamma$ = 48$^{\circ}$ and $\chi$ = -30$^{\circ}$. The Si I line is fit independently (best-fit profiles not shown in this figure).
}
\label{fig4}%
\end{figure}

To fit the photospheric Si I line, we consider two atmospheric components: a magnetic component and a field-free one. In the case of the chromospheric He I line, there are only few pixels showing multiple co-existing atmospheric components with different line-of-sight velocities. They are located around the position of X =8, Y =12 (in the coordinate of Fig.~\ref{fig2} on May 17), which is outside the filament area we are interested in. Therefore we can safely use a one-component magnetic atmosphere model to do the inversion in the scanned region, with the assumption that the magnetic field fills the resolution element completely due to its strong expansion with height \citep[e.g.][]{solanki90}. Typical observed and best-fit full Stokes profiles of the He I line are shown in Fig.~\ref{fig4} taken from the He I absorption area in the section S1 (right panels) and S2 (left panels) of the filament, respectively. Note that in the left panels the signal is well above the noise in all four Stokes profiles. Stokes $Q$ and $U$ are below the noise in the right panels.

The He I triplet can display the signature of the Hanle effect, so that in principle the inversion should take the influence of the Hanle effect on the Stokes profiles into account. However, we found that the He I Stokes $Q$ and $U$ profiles from the strong helium absorption feature exhibit the clear signatures of the Zeeman effect with no trace of the Hanle effect can be seen (an example in the left panels of Fig.~\ref{fig4}). This is consistent with the results of \cite{truj07} and \cite{cen09}, who found that in low-lying optically thick plasma structures, such as those of active region filaments, the amount of atomic-level polarization may turn out to be negligible and the emergent
linear polarization of the He I triplet in such structures be dominated by the contribution of the transverse Zeeman effect. Therefore, a Zeeman-based ME inversion of on-disk measurements of the He I triplet can retrieve reliable values for the magnetic field. Consequently, the Hanle effect is neglected for the He I triplet inversion in this work.

\section{Vector Magnetic Fields}
Figures \ref{fig5} and \ref{fig6} display the inferred magnetic field vectors in the line-of-sight (LOS) frame on May 17 in the photosphere (left column) and upper chromosphere (right column). The images in the infrared continuum at 1083.25 nm and in the He I line core are presented in panels Fig.~\ref{fig5} \textbf{I} and Fig.~\ref{fig5} \textbf{II}, respectively. It is clearly seen in the continuum image that a large penumbra-like structure (which, however, appears not to be a regular penumbra attached to a particular sunspot) was formed roughly along the netural line. Such a structure is indicative for highly inclined magnetic fields at the solar surface. The section S2 of the filament seen in H$\alpha$ overlies this penumbra-like region. At the formation height of the He I line, however, only individual absorption patches are observed above this penumbra-like structure.

\subsection{Magnetic field strength}
\begin{figure}
\includegraphics[width= 8 cm]{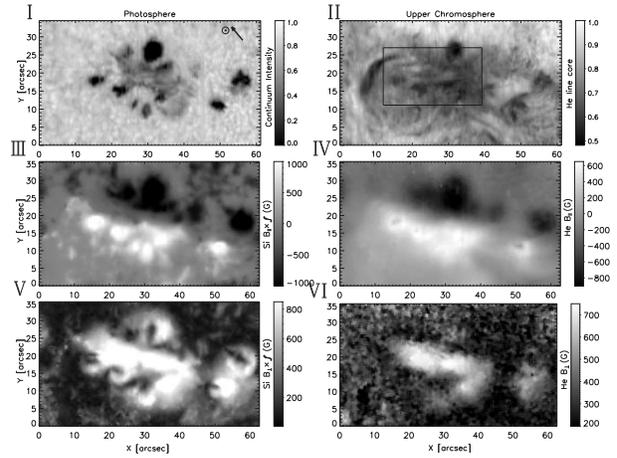}
\caption{Retrieved magnetic field information in the LOS frame of the part of active region NOAA 10763 scanned by TIP II on May 17 in the photosphere (left column)
and upper chromosphere (right column). Panel \textbf{I} shows the infrared continuum image at 1083.25 nm.
Panel \textbf{II} displays the intensity around the He I line core integrated from 1083.0 nm to 1083.06 nm. Panels \textbf{III} and \textbf{IV} depict the LOS magnetic field  in the photosphere and chromosphere, respectively. Panels \textbf{V} and \textbf{VI} show the absolute value of transverse magnetic field.}
\label{fig5}%
\end{figure}

The LOS component of the magnetic field ($B_{//}$) in the photosphere and chromosphere are shown in panels Fig.~\ref{fig5} \textbf{III} and Fig.~\ref{fig5} \textbf{IV}, respectively (in panel Fig.~\ref{fig5} \textbf{III}, we actually plot $fB_{//}$, where $f$ is the filling factor). As expected, the chromospheric magnetic field strength map is much more diffuse,
but does show some structure, partially due to the underlying sunspots. At the present resolution (about 1$\arcsec$-2$\arcsec$), the polarity distribution,
particularly the position of the boundary between the opposite polarities, is very similar in both layers. In panels Fig.~\ref{fig5} \textbf{V} and Fig.~\ref{fig5} \textbf{VI} we display the absolute value of the transverse magnetic field $B_{\bot}=\sqrt{(B_{x}^{2}+B_{y}^{2}})$ (in panel \textbf{V} it is $fB_{\bot}$). Strong transverse magnetic fields are found in and around the filament section S2 in both layers, with $B_{\bot}$ as high as 500 - 700 G in the upper chromosphere (e.g. $B_{\bot}$ $>$ 600 G is obtained inside the filament and outside a spot) and $fB_{\bot}\approx$ 600 - 800 G in the photosphere (with filling factors larger than $75\%$). In contrast, no strong magnetic fields, either in the LOS or the transverse direction, are found around the filament section S1 in both layers.

\subsection{Magnetic field orientation}
\begin{figure}
\includegraphics[width= 8 cm]{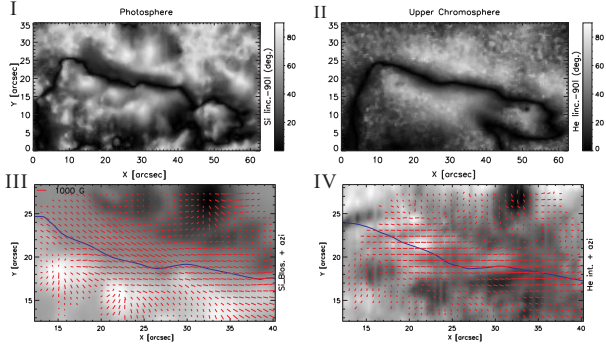}
\caption{Retrieved magnetic field information in the LOS frame in the photosphere (left column) and upper chromosphere (right column). Panels \textbf{I} and \textbf{II} provide the absolute value of the inclination of the magnetic field $|\gamma-90^{\degr}|$, where $\gamma$ is the angle between the magnetic vector and the LOS. Dark (white) shading indicates magnetic field directed perpendicular (parallel) to the LOS. The boxed area in Fig.~\ref{fig5} \textbf{II} is enlarged and plotted in \textbf{III} and \textbf{IV}, for multiple variables. \textbf{III}: transverse magnetic fields inferred from the Si I line (red bars) overplotted on the LOS magnetic field map of the photosphere. \textbf{IV}: transverse magnetic fields inferred from the He I line (red bars) superposed on the intensity image of He I line core. Lengths of the red bars indicate the strengths of the transverse magnetic fields, the directions indicate the field's orientation. Blue lines are the polarity inversion line (PIL) calculated from the smoothed LOS magnetic field in each atmospheric layer.
}
\label{fig6}%
\end{figure}

 Panels \textbf{I} and \textbf{II} in Fig.~\ref{fig6} show the absolute value of $|\gamma-90^{\degr}|$ ($\gamma$ is the magnetic field inclination angle with respective to the LOS) in the photosphere and chromosphere, respectively. Black (white) shading represents magnetic field vectors transverse (parallel) to the LOS. It is seen that the field inclination is more symmetric around the polarity-inversion-line (PIL) in the upper chromosphere than in the photosphere, where sunspots strongly structure the inclination and field strength maps. In addition, comparison of Fig.~\ref{fig5} \textbf{III} with Fig.~\ref{fig5} \textbf{IV} and of Fig.~\ref{fig6} \textbf{I} with Fig.~\ref{fig6} \textbf{II} shows that the region around filament section S2, where the magnetic fields are nearly transverse ($|\gamma-90^{\degr}|<30^{\degr}$) and the magnetic field strengths are as low as 200 G in both photosphere and chromosphere, has different size in these two layers. It is about 7$\arcsec$ wide in the photosphere and about 4$\arcsec$ in the upper chromosphere. It implies that the opposite polarities are located closer to each other in the chromosphere than in the photosphere.

Magnetic field azimuth angles in the two layers are displayed in panels \textbf{III} and \textbf{IV} of Figure~\ref{fig6}. Only a part of the scanned region, inside the box in Fig.~\ref{fig5} \textbf{II}, is plotted, since this is the region of strongest interest for the structure of the filament. The transverse magnetic field inferred from the Si I line is over-plotted onto the LOS magnetic field map obtained from the same line in Fig.~\ref{fig6} \textbf{III}. In Fig.~\ref{fig6} \textbf{IV}, the chromospheric transverse magnetic field derived from the He I line is superposed on the intensity image of the He I line center. At first sight, the orientations of transverse fields (indicated by the orientation of the bars) in both layers generally follow the long axis of the filament, i.e. they are roughly tangential to the PIL (indicated by blue solid lines in Fig.~\ref{fig6} \textbf{III} and Fig.~\ref{fig6} \textbf{IV}). A closer inspection, however reveals some difference between the two layers. Whereas there is a good alignment between the magnetic field azimuth angle and the PIL in the photosphere, the azimuth angle is tilted by 20$^{\circ}$ - 30$^{\circ}$ to the PIL in the chromosphere. Consequently, the field azimuth follows the segmented absorption features seen in the He I line, i.e. those features are field-aligned.

\subsection{Disturbance of LOS magnetic signals and LOS velocities }
\begin{figure}
\centering
\includegraphics[width= 7 cm]{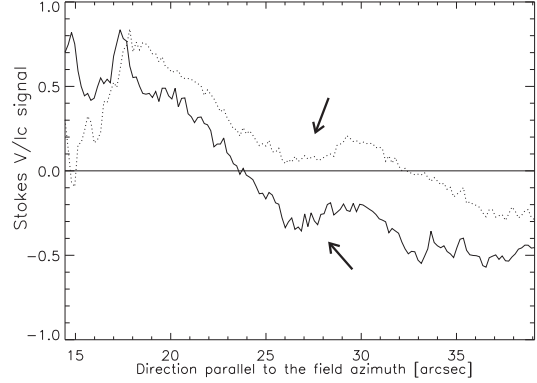}
\caption{Variation of spectrally integrated He I $V/I_{c}$ along the magnetic azimuth direction. Wavelength integration is over a range of 0.4 \AA in the blue wing of the He I line. The plotted cuts lie along Y = 20 (solid line) and Y = 18 (dotted line) in the coordinates of Fig.~\ref{fig6}. The arrows point to the S-shaped disturbance of the signal. The horizontal line at $V/I_c  = 0$ indicates polarity inversion.
}
\label{fig7}%
\end{figure}

Next we carefully investigate the magnetic field in the vicinity of the PIL (i.e. the region with $|\gamma-90^{\circ}|<30^{\circ}$) in both studied atmosphere layers. The retrieved inclination maps (in Fig.~\ref{fig5}, panles \textbf{III} and \textbf{IV}) display a simple polarity inversion line, with the two dominant opposite polarities on either side. However, the He I Stokes $V$ signal (i.e., the integrated Stokes $V/I_{c}$ within a wavelength range of 0.4 \AA\ in the blue wing of the He I line) displays a disturbance along the field line, i.e. when moving in a direction parallel to the azimuth angle. As shown in Fig.~\ref{fig7}, superposed on the general trend and sign change when going from left to right, close to the PIL the Stokes $V$ signal displays a fluctuation, producing an S-shape. Along the S-shape the field lines are horizontal within 5$^{\circ}$ - 10 $^{\circ}$, which is smaller than the local heliocentric angle of 16$^{\circ}$ ($\mu = cos\theta \approx 0.96$). Such a signal is indicative of a dip near the top of a magnetic loop or flux rope. In an entirely symmetric loop or flux rope, a dip in the field lines at the apex would produce an S-shaped distribution consistently on the limbward side of the PIL. As can be deduced from Fig.~\ref{fig7}, the S-shaped distribution can lie on either side of the PIL, depending on where it is cut. This indicates that the dip is not always at the apex and probably the flux rope is somewhat skewed.

\begin{figure}
\centering
\includegraphics[width= 8 cm]{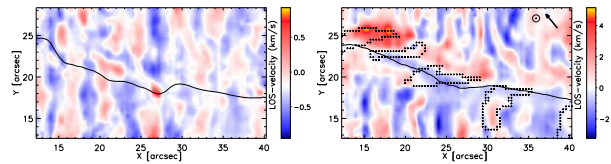}
\caption{LOS velocities derived from the Stokes I profiles of the Si I line (left panel) and the He I line (right panel) inside the region enclosed by a rectangle in panel Fig.~\ref{fig5} \textbf{II}. The solid line is the PIL calculated from the LOS magnetic field of each layer. Segmented He I filaments are outlined by dotted contours in the right panel. The direction towards disk center is indicated by an arrows.
}
\label{fig8}%
\end{figure}

LOS velocities obtained from the inversions in both the photosphere and chromosphere are illustrated in Fig.~\ref{fig8}. There is no consistent flow in the photosphere associated with the filament, but even though the velocities are not larger, there clearly is an extended upflow along the PIL in the chromosphere. This upflow is also present inside the segmented He filament and is flanked by downflows.

\subsection{180 degree ambiguity removal}
\begin{figure}
\centering
\includegraphics[width= 8 cm]{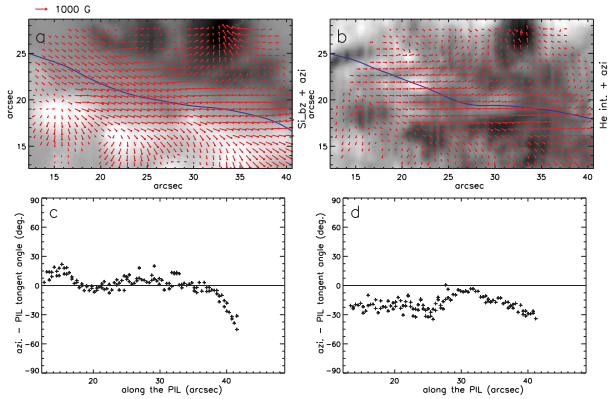}
\caption{Magnetic fields inside the region delimited by a rectangle in panel \textbf{II} of Fig.~\ref{fig5}, after the 180$^{\circ}$ disambiguation of the azimuth angle and transformation to the local solar frame. \textbf{a}: Photospheric horizontal magnetic fields inferred from the Si I line (arrows)
overplotted on the photospheric vertical magnetic field map. \textbf{b}: Horizontal magnetic fields inferred from the He I line superposed on the intensity
image of the He I line core. The arrows in panels \textbf{a} and \textbf{b} indicate the orientations and strengths of the horizontal magnetic fields. The blue line is the PIL calculated from the smoothed vertical magnetic fields of each layer. \textbf{c-d}: The acute angle between the orientation of the horizontal field and the tangential direction of the PIL in the photosphere (\textbf{c}) and chromosphere (\textbf{d}). A positive sign of the angle means an angle directed from negative to positive polarity.
}
\label{fig9}%
\end{figure}

We apply the non-potential magnetic field calculation method, NPFC \citep{georgoulis05}, to resolve the 180$^{\circ}$ ambiguity on the magnetic field azimuth angle and transfer the magnetic field vector into the local solar frame. We illustrate the resulting projection of the magnetic field vector on the solar surface for section S2 of the filament in Fig.~\ref{fig9}\textbf{a} (photosphere) and Fig.~\ref{fig9}\textbf{b} (chromosphere), which  are very similar to those panels \textbf{III} and \textbf{IV} of Fig.~\ref{fig6} (because $\mu$ is close to unity), except that now arrows indicate the direction of the horizontal component of the field. The PIL moves slightly northward when transferring into the local solar frame, which is consistent with the frame transformation since the AR filament is located in the southern hemisphere. Along the PIL (strictly speaking, where the angle of the field lines to the local solar surface is smaller than 3$^{\circ}$), we check the acute angle between the tangential direction of the PIL and the orientation of the adjacent horizontal magnetic field. In the photosphere the angles are on average smaller than 10$^{\circ}$ with a tendency towards positive values except near one end (see Fig.~\ref{fig9}\textbf{c}). In the chromosphere the angles are on average around 20$^{\circ}$ and always negative (see Fig.~\ref{fig9}\textbf{d}). The negative sign of the acute angle in the upper chromosphere implies a horizontal field directed from positive to negative magnetic polarity, which corresponds to the normal magnetic field configuration. Hence we clearly have a normal configuration in the upper chromosphere sampled by the He I triplet and a hint of a concave topology in the photosphere.

\section{Discussions}
In this paper, we have presented spectro-polarimetric observations of an AR filament, which we analyzed together with other space- and ground-based observations. The inversion code HeLix$^{+}$, based on the Milne-Eddington approximation, is used to retrieve the magnetic field vectors in photosphere and upper chromosphere, respectively, from simultaneously observed Stokes profiles in the Si I 1082.7 nm and He I 1083.0 nm lines.
The AR filament comprises two sections, a curved section S1 and a roughly E-W directed section S2. We inferred transverse magnetic fields reaching 500 - 700 G in the upper chromosphere of section S2. This confirms the findings of \cite{kuckein09} for another AR filament and shows that such strong fields are not uncommon among AR filaments. In section S1 of the filament, however, the field strength was below 200G everywhere, suggesting that fields above 500 G are not always found in AR filament.

In the following we summarize and discuss our main findings, particularly for section S2 of the studied AR filament.

\begin{enumerate}
  \item MDI Magnetograms obtained during May 15 to May 17 show that the magnetic flux in the region of the filament increases until May 16 and then decreases again, with the changes in the flux taking place mainly in situ. The opposite magnetic polarities in the photosphere on both sides of the filament converge toward each other, mainly during the time when the magnetic flux is decreasing. As pointed by \cite{lit10} such an in situ decrease or "cancellation" may or may not imply the presence of reconnection. The observed magnetic flux evolution may also result from the emergence of the lower reaches of a flux rope that is buoyantly rising into the atmosphere (\cite{oka08}; see the qualitative comparison between the reconnection scenario and the emerging flux rope scenario in \cite{wang07}).

  \item The filament (S2) was first seen in H$\alpha$ on May 16, which roughly coincides with the time that the magnetic flux starts to decrease again. We do not know if a filament or small, segmented filaments were visible in the He I line on May 16. The AR filament is visible in both H$\alpha$ and He I lines on May 17, but it exhibits different morphologies in maps recorded in these two lines. The filament seen in the H$\alpha$ line core is complete, while in He I it is partial, i.e. it is composed of elongated field-aligned features restricted to a few locations.

  \item The opposite magnetic polarities in the chromosphere are closer to each other than in the photosphere.

  \item In the photosphere, the magnetic azimuth angle is nearly aligned along the PIL, while in chromosphere, the azimuth angle crosses the PIL at an angle of 20$^{\circ}$ - 30$^{\circ}$. In addition, in the corona the EUV loops are observed to be more perpendicular to the PIL. These twists can be interpreted in term the magnetic field in and around the filament exhibiting different degrees of twist at different atmospheric layers, which is consistent with twisted flux tube or flux rope models of filaments (e.g.  Fig.~5 of \cite{priest89}, Fig.~7 of \cite{aulanier02}). It is possible, however, that coronal loops are not part of the flux rope itself, but part of the overlying arcade holding it down.

  \item The LOS magnetic field signal displays the signal of a dip in the magnetic field close to the PIL. The dip must be shallower than 16$^{\circ}$ (the local solar heliocentric angle), so that it does not cause a polarity inversion.

  \item In the chromosphere, small but consistent upflows are found along the PIL and inside the segmented He filaments. Downflows are found at its sides. Such a flow structure is typical of slowly rising emerging loop-like structures \citep[e.g.][]{solanki03,xu10}. It illustrates that this filament is forming (it was still incomplete during the scan) and material is slowly rising together with the field along the center of the filament, but flowing down its sides.

  \item After the 180$^{\circ}$ ambiguity is resolved for the field vector in both layers, the direction of the horizontal component of the magnetic field in the photosphere mainly points along the PIL, but with a slight preference of going  from negative polarity to positive polarity, \textbf{which indicates a concave structure,} although with very little twist in the flux rope at this atmospheric layer. In contrast, horizontal magnetic fields in the upper chromosphere are found directed from positive polarity to negative polarity, corresponding to a normal magnetic field configuration. The angle between field azimuth and PIL has increased to 20$^{\circ}$ - 30$^{\circ}$ in the chromosphere. It is interpreted in terms of a flux, this implies a much larger twist in this layer. The deduced configuration is in agreement with the investigation of \cite{bom94} on prominences in the vicinity of active regions. Using the He I D$_3$ line, \cite{bom94} found that the two prominences in their samples close to active regions are in normal configuration.
\end{enumerate}

\begin{figure}
\includegraphics[width= 6 cm]{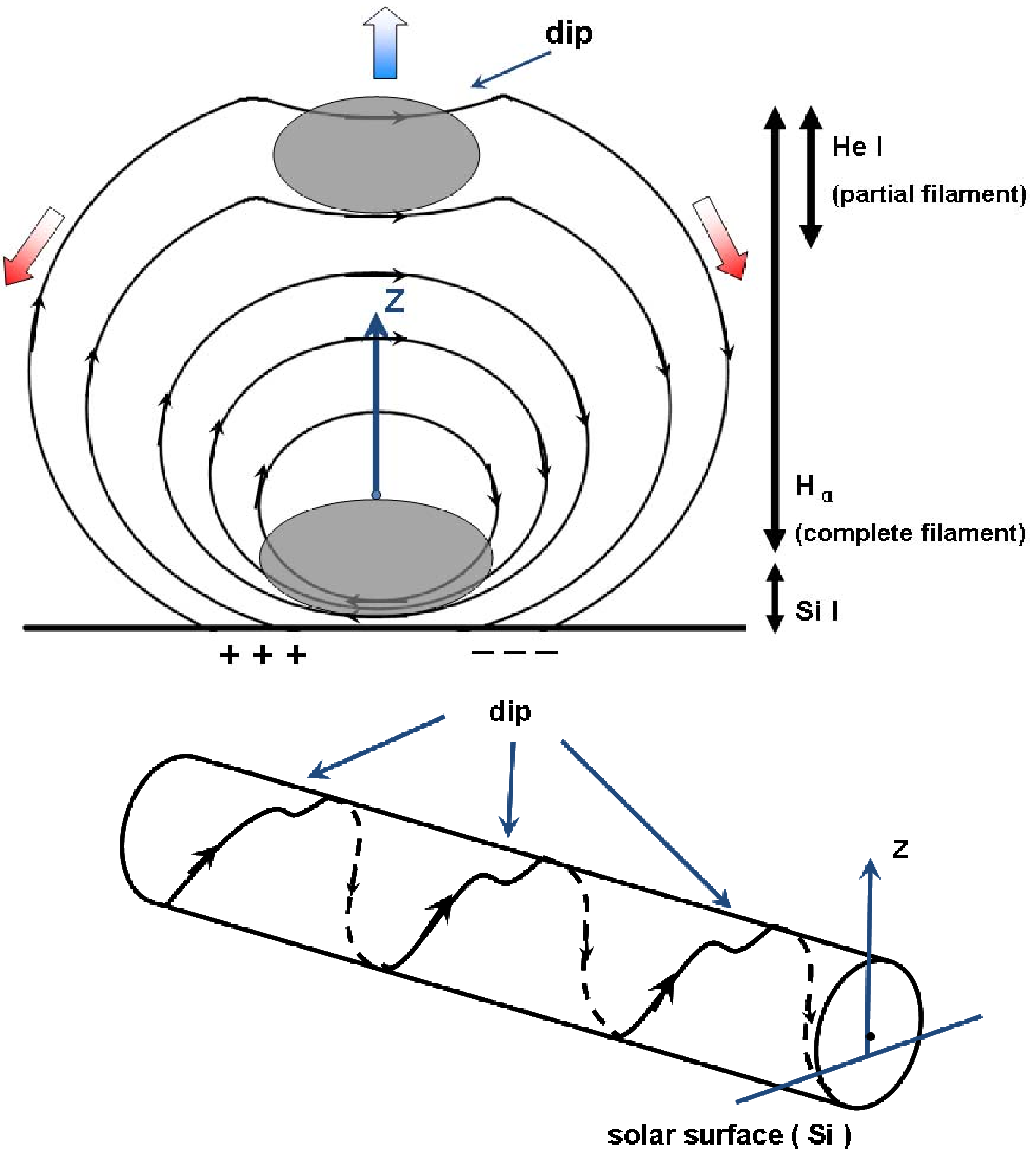}
\caption{Sketch of the proposed scenario based on the observational facts gathered at two atmospheric layers. The filament is defined by a magnetic flux rope.
Upper panel: Cross-section of the flux rope, as viewed along the long axis of the flux rope. Filled grey ovals represent the filament material (see the text).
There is a flat dip in the magnetic field lines near the top of the flux rope in order to support the He filament material. The thick blue arrow indicates upflows in the upper part of the flux rope, produced by a gradual rise in the magnetic structure. Two thick red arrows represent downflows along the field lines at its sides. The thick horizontal line represents the solar surface. The heights sampled by the Si I, H$\alpha$ and He I lines are marked by thick vertical arrows, respectively. Lower panel: A perspective view of the flux rope. A representative magnetic field line on the surface of the flux rope is indicated by curved lines. Solid (dashed) lines represent the field line on the front (rear) side of the flux rope.
}
\label{fig10}%
\end{figure}

Considering all the points above and under the assumption that the 180$^{\circ}$ ambiguity removal gives the correct direction of the magnetic vector, we put forward a scenario for the structure of the investigated filament, which is illustrated in Fig.~\ref{fig10}. We describe the magnetic structure underlying the filament by a helical flux rope, whose cross-section and side views are both sketched. An important characteristic of the proposed configuration is that beside the natural and stable locations produced by the flux rope where material can be suspended (troughs of the helical windings visible in the lower part of the flux rope), there are dips in the field lines in the upper part of the rope. These dips are rather flat and produce the disturbance of the LOS magnetic signal visible in Fig.~\ref{fig7}. Consequently, the material can be stored at two heights. The different height ranges in the solar atmosphere sampled by Si I, H$\alpha$ and He I lines are marked in the upper panel. The formation of the He I triplet is restricted to the upper part of the chromosphere, to which the ionizing EUV radiation (below 50.4nm) from the corona can penetrate The absence of asymmetries in the He I absorption profiles suggests that the He line is formed in a narrow atmospheric layer, consistent with an ionization process dominated by EUV radiation). The H$\alpha$ formation layer is significantly broader, covering essentially the whole chromosphere and overlapping with the He I triplet formation layer.

As a result, only the upper part of filament material is visible in He I since the He I triplet is only formed in the upper layers. However, both the top and deep parts of the filament material can be seen in H$\alpha$. This explains why the observed filament has different morphologies in these two lines. In addition, the layer sampled by the Si I line is located \emph{beneath} and \emph{close} to the center of the axis of the flux rope, while the layer probed by the He I line is located \emph{above} and\emph{ further away} from the axis. Therefore the winding field lines below the axis of the helix exhibit a concave, but are tilted only very sightly with respect to the PIL at the heights sampled by Si I (Twist is expected to be small close to the axis of the flux rope). The winding field lines above the axis show normal configuration, tilted more strongly with respect to the PIL at the height sampled by He I. In this geometry, the opposite polarities in the chromosphere are closer to each other than in the photosphere, also in agreement with the observations. Note, however, that the sketch in Fig.~\ref{fig10} is simplified. In reality the flux rope must be asymmetric, with a changing asymmetry over its length, since the signature of the dip is formed on different sides of the PIL at different positions along its length. Weak but consistent flows found in the chromosphere also fit into this picture. The upflows are seen at the top of the rope perpendicular to the field lines, while the downflows on the sides of the filament detected by He I probably follow the field lines, which is consistent with an emerging flux rope.

This picture is also supported by the evolution of the magnetic flux, with initially increasing flux (rising flux rope, with axis still below solar surface), followed by decreasing flux and converging polarities (axis of the flux rope is now above the solar surface). This second phase coincides with the formation of the filament visible in H$\alpha$ (which also indicates that the filament visible in H$\alpha$ is mainly located below the flue rope axis). Given this evolution of the filament, we further infer that, unlike the emergence event studied by \cite{oka08}, i.e., new flux emerging under an existing filament, we are seeing the emergence of a flux rope that is producing a filament as it emerges.

Although this scenario rather nicely explains our observational findings, we cannot rule out alternative scenarios, with the difference between the filament seen in H$\alpha$ and He I being due to, e.g., differences in the optical depths of the two lines. Observations of the further evolution of this or similar AR filaments could cast additional light, for instance, by probing whether the magnetic field in the chromosphere changes from a normal configuration to a concave topology as the flux rope emerges further, i.e., when the height sampled by He I lies beneath the flux rope axis, so that He I 1083.0 nm and H$\alpha$ sampled the same filament. However due to the paucity of high quality spectro-polarimetric observations at later times, and multiple flares occurring near the location of the filament after the analyzed observation, not much more can be said at present.

\section{Conclusion}
We have investigated the vector magnetic field measured simultaneously in the photosphere and upper chromosphere ---
from spectro-polarimetric scans made in the Si I 1082.7 nm and He I 1083.0 nm line --- of an AR filament before its flare associated activation and partial eruption.
At the time of observation, this filament appeared to be lying low enough in the atmosphere, so that we could observe characteristic signatures of a strong-field flux rope by combining the vector magnetic fields of both atmosphere layers. Thus, we have learned that strong magnetic fields, with a retrieved field strength of 600 - 800 G in the He I line formation height, are not uncommon among AR filaments. Considering all the observational facts and the assumption that the 180$^{\circ}$ ambiguity is correctly resolved in this filament area, we propose a scenario of the magnetic field structures associated with the AR filament and conclude that we are observing the emergence of a flux rope that is producing a filament as it emerges. As an unusual feature of this structure we find that there are probably two filaments overlying each other, one is in the lower, one in the upper part of the flux rope, i.e. one displaying normal magnetic configuration and the other a concave topology. We speculate that the upper (partial) filament is not stable and eventually drains back into  the sun in the course of the evolution. We can not completely rule out other configurations, however, due to a lack of magnetic vector information in the chromosphere and insufficient knowledge of the magnetic field's evolution. Observations of further AR filaments are of great interest to test the scenario proposed here.

\begin{acknowledgements}
The authors wish to thank B. Schmieder for fruitful discussion. The data used in this paper were obtained with the German Vacuum Tower Telescope in the Teide
Observatory of Spain. This work was partly supported by the National Natural Science Foundation grant No. 10933003 and 11103075 of China and partly by the WCU grant
 No. R31-10016 funded by the Korean Ministry of Education, Science and Technology.

\end{acknowledgements}

\clearpage
\end{document}